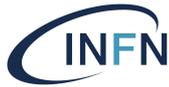

ISTITUTO NAZIONALE DI FISICA NUCLEARE

# PHYSICALLY IMPOSSIBLE?

D. Domenici, G. Mazzitelli

*INFN - Italian Institute for Nuclear Physics (ITALY)*

## Abstract

Halfway between the experiment and the focus group, between the quiz and a game, we have experienced a new format to "focus" on sustainability and the fundamental laws of thermodynamics and its principles. Concepts as reversibility, efficiency and entropy, are then "visualized" by the participants, showing the relations with the economic value, waste, the energetics budget and raw material costs are explained from a different point of view, proving the physical limits to the economic growth and the environmental exploitation.

Keywords: thermodynamics, physics teaching methodology, research and education, researchers at school.

# 1 INTRODUCTION

As researchers working at the Italian National Institute for Nuclear Physics (INFN) [1], the outreach is part of our mission [2]. We often go in high schools to make seminars introducing cosmology, high energy physics, quantum mechanics and general relativity, sometimes testing new approaches to science communication [3]. Since 2011, after the Fukushima Daiichi nuclear disaster, many high school teachers have started to ask us to speak about nuclear energy and more in general of the energetic issue.

After some time spent making boring presentations, we have started a different approach: the exploitation of energy is strongly connected to the issue of sustainability, the small subset where environment, economics and societies meets. But, while the rules of the society and of the economy are made by humans, the environmental ones are dictated by nature that follows the principles of thermodynamics which set the limits. That means that we can make a better introduction to thermodynamics by investigating the society and the economy. This pushed us to exploit one of the most common methodologies to understand societal and economic issues: the focus group.

# 2 METHODOLOGY

High school's students participating in our focus group are typically divided into groups of about 5 persons, equipped with a post-it block and a pen. We start proposing few relevant societal and physical keywords about sustainability, ecology and energy, and challenging the groups to brainstorm about and to write-down words and concepts that are eventually presented to all the audience.

We act as moderator of the discussion alternating the brainstorming with classical slides showing plots and diagrams to introduce and analyze the various concepts and reveal the underlying physical principles.

# 3 RESULTS

The method has been successfully tested at high schools and public science events, generating interesting feedback. The following plots present the preliminary results collected in 3 years of survey in the schools. The survey was a volunteer, typically attended by 20/30% of students.

*Table 1. Age distribution of the participants to the survey.*

| Age | < 10 | 16 | 17 | 18 | 19 | > 30 |
|---|---|---|---|---|---|---|
| Number of participants | 1 | 6 | 52 | 21 | 3 | 1 |

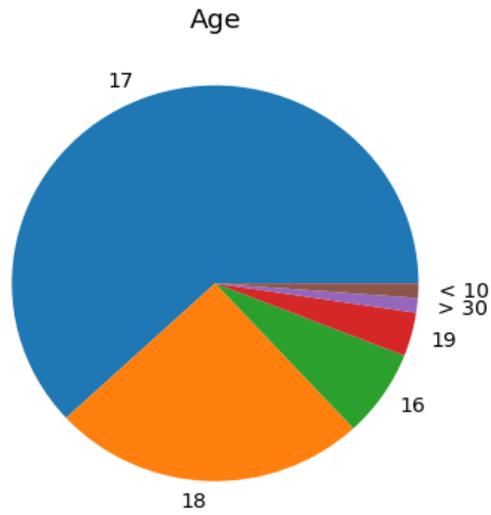

*Figure 1. Age distribution of the participants to the survey.*

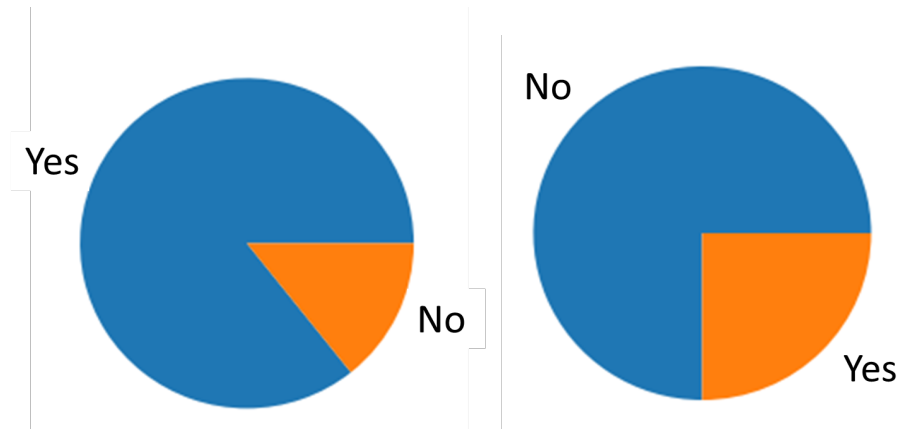

*Figure 2. Answers to the questions "are you interested in science?" (left) and "do you want to be a scientist?" (right).*

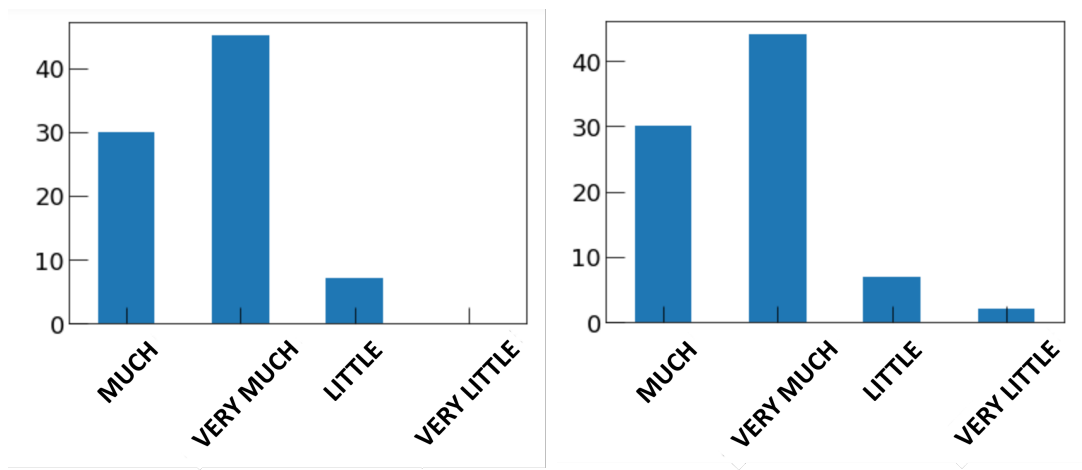

*Figure 3. Answers to the questions "how much did you find the lesson interesting?" (left) and "how much was the lesson adapted to your preparation?" (right).*

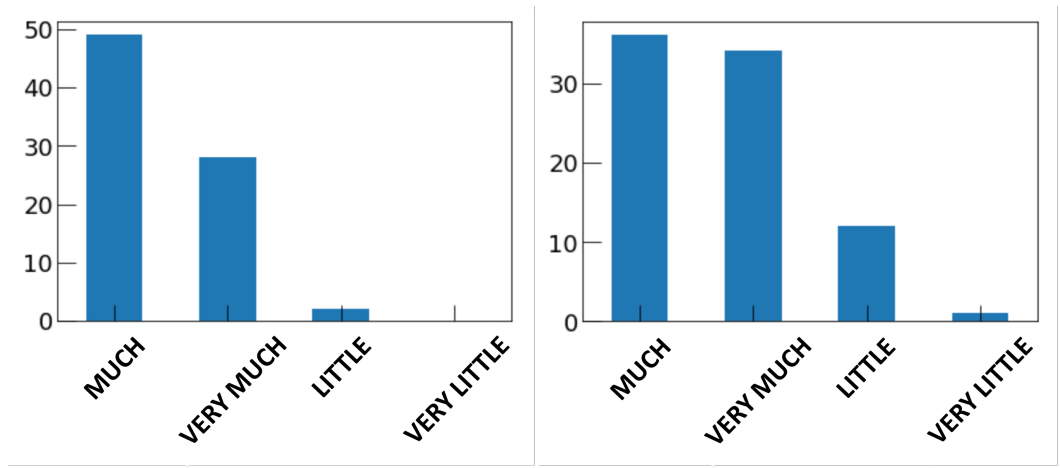

*Figure 4. Answers to the questions "how clear was the teacher?" (left) and "how adequate were the materials and exercises presented?" (right).*

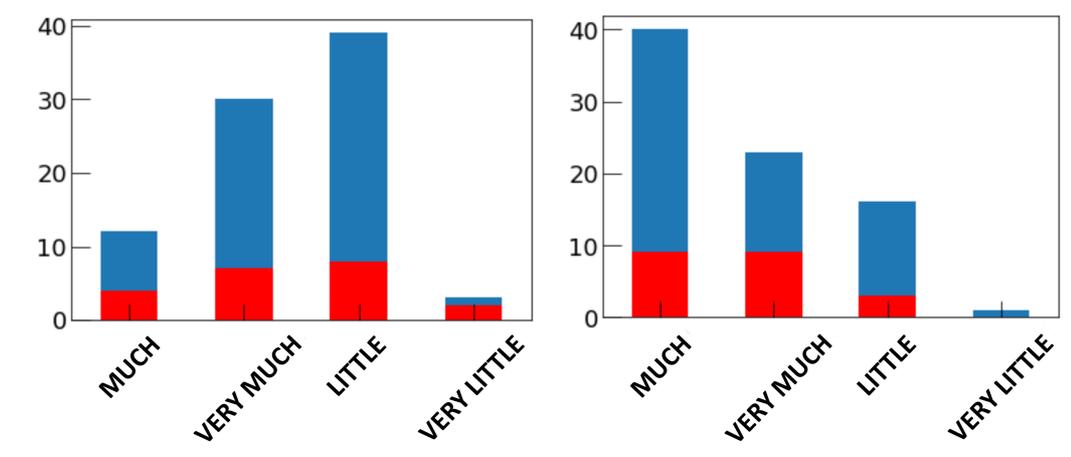

*Figure 5. Answers to the questions "how much do you think the lesson will be useful for your studies?" (left) and "how important do you think the topic is for your future?" (right).*

The analysis of the survey shows that students were in general very satisfied after the participation in the focus group. It is worth remarking that most of the students say to be interested in science although most people do not intend to be a scientist (see Figure 2). Science is therefore recognized as an important aspect of life whatever our own job is. Moreover, there are no strong differences among students that would like to be a scientist (figure 5 - red) respect to others in the last two questions that try to evaluate the short and future impact of the lesson.

Besides a very positive feedback about the participant's satisfaction and the teacher's preparation (Figure 3 and Figure 4), there is an interesting trend shown in Figure 5: the topic presented is not fully identified as a standard part of the school's program, even though concepts like energy and entropy are discussed. This means that both the language and the methodology used result different with respect to those typical in their classrooms. On the other hand, the topic is fully identified as necessary to know to be a responsible citizen.

## 4   CONCLUSIONS

We present our experience as physics researchers in holding a focus group by high schools and science festivals. Instead of using the "definition to application" path typical of school programs, we tried a reverse "application to definition" way that start from general social concepts, like "sustainability", "ecology", "recycling", to eventually show that behind all of those there are precise physics quantities, like "energy", "efficiency" and "entropy", related by precise physics laws.

At the end of the game, people seem to have better understood the laws of thermodynamics!

**ACKNOWLEDGEMENTS**

This project is part of the initiatives promoted by the researchers of the Frascati Laboratory of the National Italian Institute of Nuclear Physics (INFN) to promote research and involve youth in science.